\documentclass[%
print,
 amsmath,amssymb,
 aps,
]{revtex4}

\usepackage[sc]{mathpazo}
\usepackage{graphicx}
\usepackage{dcolumn}
\usepackage{bm}
\usepackage{dsfont}
\usepackage[landscape,papersize={297.1mm,210mm},left=1.9cm,right=1.6cm,top=2.7cm,bottom=2.8cm]{geometry}
\usepackage[                   
            pdfstartview=FitH,
            colorlinks, 
            pdfborder=001,   
            linkcolor=blue,
            anchorcolor=blue,
            citecolor=blue,
            urlcolor=blue
            ]{hyperref}
\usepackage{amsthm}
\makeatletter

\newcommand{\Rmnum}[1]{\expandafter\@slowromancap\romannumeral #1@}
\makeatother

\begin{document}
\title{Quantum properties in the four-node network}

\author{Yanwen Liang$^1$}
\affiliation{College of Physics, Hebei Key Laboratory of Photophysics Research and Application, Hebei Normal University, Shijiazhuang 050024, China}
\author{Fengli Yan$^1$}
\email{flyan@hebtu.edu.cn}
\affiliation{College of Physics, Hebei Key Laboratory of Photophysics Research and Application, Hebei Normal University, Shijiazhuang 050024, China}

\author{Ting Gao$^2$}
\email{gaoting@hebtu.edu.cn}
\affiliation{School of Mathematical Sciences, Hebei Normal University, Shijiazhuang 050024, China}

\begin{abstract}
There are different preparable quantum states in different network structures. The four nodes as a whole has two situations: one is the four nodes in a plane, the other is the  four nodes in the space. In this paper, we obtain some properties of the quantum states that can be prepared in four-node network structures. These  include the properties of entropy, entanglement measure, rank and multipartite entangled states. These properties also mean that the network structures impose some constraints on the states that can be prepared in a four-node quantum network. In order to obtain these properties we also define  $n$-partite mutual information of the quantum system, which satisfies  symmetry requirement.\\

\end{abstract}

\pacs{ 03.67.Mn, 03.65.Ud, 03.67.-a}

\maketitle

\section{Introduction}

By using quantum network one can exchange corresponding quantum information on different nodes through quantum channels, such as quantum entanglement \cite{1,2,3,4,5,6,7,8} and quantum key distribution \cite{15, 115}. Therefore, quantum network plays an important role in the study of quantum correlation \cite{9,10,11}. The great progress has been made in the  research and application of quantum network structure and quantum entangled states prepared in the quantum network in both theory and experiment \cite{12, 13, 14, 16, 17, 18}. Of course, the practical application of quantum network technologies may also affect the real life in the future.

In the network  system, quantum states are distributed to different nodes \cite{12,13}. It is assumed that sources are independent in the quantum network. It  means different sources are separable. However, different particles in the same source can be entangled \cite{14,15}, so there are  a lot of situations. For example, one can has the situation in which a source can only generate two particle entangled states. Of course,  there also exists the situation in which three particle entangled states can be produced by the source.

Apparently, the quantum system consisting of different number particles have many quantum network structures and a lots of different preparable states \cite{16,17}. This paper only studies several different quantum network structures of four  nodes.  The structure of four-node quantum network has two different situations: one is the network structure of four nodes in a plane, the other is the network structure of four nodes in the space.  We assume that the sources do not intersect.  Under the assumption of no intersection of sources, the four nodes in a plane have only one quadrangular network structure, while the four nodes in the space have two kinds of  network structures.  Different quantum states can be prepared in different network structures. Based on the pioneer work in three nodes \cite{18}, we will mainly discuss the   network structures  in the above three special situations, and deduce some properties and constraints of the quantum states that can be prepared.

The paper is organized  as follows. In section II we define $n$-partite mutual information  in quantum systems and obtain some properties of states prepared in a  quadrangular  network structure.  The properties of quantum states prepared in two kinds of network structures with four nodes in the space are mainly investigated in section III.  Among them, the two network structures of four nodes in the space with six sources, where each source generates  the two-partite entangled state, and four sources, where every source produces three-partite entangled states are discussed respectively. A summary is given in section IV.

\section{Four nodes in a plane}

Let us discuss the situation in which four nodes are in a plane. This network is called independent quadrangular  network  (IQN). As shown in Fig.1(a), it has four nodes, namely \emph{A}, \emph{B}, \emph{C}, and \emph{D}, which are formed by pairing of two particles in each of the four sources $\rho_{\alpha}$, $\rho_{\beta}$, $\rho_{\gamma}$, and $\rho_{\delta}$ \cite{19}. The states of two particles in each source are the two-partite  entangled states \cite{20,21,22}. $\rho_{\alpha}$, $\rho_{\beta}$, $\rho_{\gamma}$, and $\rho_{\delta}$ is shared respectively by two nodes of \emph{A}, \emph{B}, \emph{C}, and \emph{D}. It means that each of the four nodes \emph{A}, \emph{B}, \emph{C}, and \emph{D} receives two particles. Exactly, \emph{A} receives two particles respectively from the source $\rho_{\alpha}$ and $\rho_{\beta}$; \emph{B} receives the two particles respectively from the source $\rho_{\beta}$ and $\rho_{\gamma}$; \emph{C} receives two particles respectively from the source $\rho_{\gamma}$ and $\rho_{\delta}$; \emph{D} receives two particles respectively from the source $\rho_{\delta}$ and $\rho_{\alpha}$. The two particles at the each node can be applied by a local unitary matrix. The four local unitary matrices are  denoted by $U_{A}$, $U_{B}$, $U_{C}$, and $U_{D}$ respectively.

\begin{figure}[h]
\centering
\includegraphics[height=0.33\textwidth]{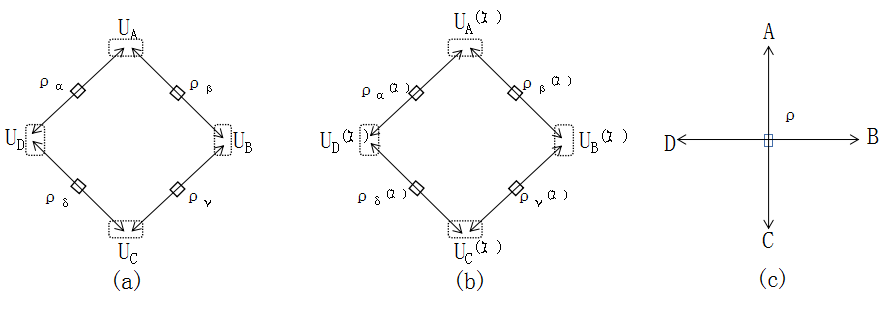}
\caption{Quadrangular network. (a) An independent quadrangular network (IQN) in which each source is a two-partite entangled state. (b) A classically correlated quadrangular network (CQN) in which every source and every node are classically correlated by sharing the random variable $\lambda$. (c) The four-partite quantum states which can not be generated in IQN and CQN. \label{figure1}}
\end{figure}

 We use $\Delta_{\mathrm{IQN}}$ to represent the set of quantum states that can be prepared in an independent quadrangular network structure shown in Fig.1(a). Obviously, the quantum state $\rho$ in the set $\Delta_{\mathrm{IQN}}$ can be written as
\begin{equation}\label{1}
\begin{aligned}
\rho=(U_{A}\otimes U_{B}\otimes U_{C}\otimes U_{D})(\rho_{\alpha}\otimes\rho_{\beta}\otimes\rho_{\gamma}\otimes\rho_{\delta})(U_{A}^{\dag}\otimes U_{B}^{\dag}\otimes U_{C}^{\dag}\otimes U_{D}^{\dag}).
\end{aligned}
\end{equation}
Here, $U_{A}$ only acts on the system consisting of a particle of $\rho_{\alpha}$ and a particle of $\rho_{\delta}$, it does not act on the other particles which are not at the node \emph{A} at same time. Similarly, other unitary matrices cannot act on the states of $\rho_{\alpha}$, $\rho_{\beta}$, $\rho_{\gamma}$, and $\rho_{\delta}$ at same time.

Next we will discuss Von Neumann entropy of the quantum states in the set  $\Delta_{\mathrm{IQN}}$. In this paper, entropy refers  to Von Neumann entropy.

Because entropy $S(\rho)$ of quantum state $\rho$ is invariant under unitary transformation, and additive on tensor products, for the quantum states shown by Eq.(1), we have
$$S(\rho)=S(\rho_{\alpha})+S(\rho_{\beta})+S(\rho_{\gamma})+S(\rho_{\delta}).$$
The three-partite entropy is extended to
$$S(ABC)=
S(\mathrm{tr}_{D}\rho_{\alpha})+S(\rho_{\beta})+S(\rho_{\gamma})+S(\mathrm{tr}_{D}\rho_{\delta}),$$
the two-partite entropy is extended to
$$S(AB)=
S(\mathrm{tr}_{D}\rho_{\alpha})+S(\rho_{\beta})+S(\mathrm{tr}_{C}\rho_{\gamma}),$$
the one-partite entropy is extended to
$$S(A)=
S(\mathrm{tr}_{D}\rho_{\alpha})+S(\mathrm{tr}_{B}\rho_{\beta}),$$
and similarly for others.

Based on entropy, one can study the mutual information of the quantum system.  The two-partite quantum  mutual information of a quantum system composed of \emph{X} and \emph{Y} is defined as \begin{equation}\label{I2}
I_{2}(X:Y)=S(X)-S(X|Y)=S(X)+S(Y)-S(XY),
\end{equation}
where $S(X|Y)$  stands for conditional entropy and $S(XY)$ is the joint entropy. Three-partite quantum  mutual information measuring the common information of subsystems \emph{X}, \emph{Y}, and \emph{Z}, was defined as \cite{18}
\begin{equation}\label{I3}
I_{3}(X:Y:Z)=I_{2}(X:Y)+I_{2}(X:Z)-I_{2}(X:YZ)=S(X)+S(Y)+S(Z)+S(XYZ)-S(XY)-S(XZ)-S(YZ).
\end{equation}

We define four-partite quantum mutual information of the quantum system composed of \emph{T}, \emph{X}, \emph{Y}, and \emph{Z} as
\begin{equation}\label{2}
\begin{aligned}
I_{4}(T:X:Y:Z)=I_{3}(T:X:Y)+I_{3}(T:X:Z)-I_{3}(T:X:YZ).
\end{aligned}
\end{equation}
By substituting the formula of the three-partite quantum   mutual information into the four-partite quantum  mutual information, one can obtain
\begin{equation}\label{3}
\begin{aligned}
I_{4}(A:B:C:D)=&S(A)+S(B)+S(C)+S(D)+S(ABC)+S(ABD)+S(BCD)+S(ACD)\\
&-S(AB)-S(AC)-S(AD)-S(BC)-S(BD)-S(CD)-S(ABCD)\\
=&x_{1}-x_{2},
\end{aligned}
\end{equation}
where
$$x_{1}=S(A)+S(B)+S(C)+S(D)+S(ABC)+S(ABD)+S(BCD)+S(ACD),$$
and
 $$x_{2}=S(AB)+S(AC)+S(AD)+S(BC)+S(BD)+S(CD)+S(ABCD).$$
Obviously, the four-partite quantum mutual information is invariant under arbitrary  permutation of the subsystems.

Furthermore, we define the $n$-partite quantum mutual information of the system consisting of subsystems $X_1, X_2, \cdots, X_n$. We use $\sigma_{X_1 X_2 \cdots X_n}$ to express a quantum state of the system.  We define the $n$-partite quantum mutual information of the system as
\begin{equation}
\begin{array}{ll}
I_{n}(X_1: X_2: \cdots : X_n)=&(-1)^{1-1}\sum_{i=1}^n S(\sigma_{X_i})+(-1)^{2-1}\sum_{1\leq i< j\leq n}S(\sigma_{X_iX_j})+(-1)^{3-1}\sum_{1\leq i< j<k\leq n}S(\sigma_{X_iX_jX_k})\\
&+(-1)^{4-1}\sum_{1\leq i< j<k<l\leq n}S(\sigma_{X_iX_jX_kX_l})+\cdots+(-1)^{n-1}S(\sigma_{X_1X_2\cdots X_n}),\\
\end{array}
\end{equation}
where $\sigma_{X_iX_j\cdots X_l}=\mathrm{tr}_{X_qX_r\cdots X_t}\sigma_{X_1 X_2 \cdots X_n}$ is the reduced density  matrix for the subsystems $X_i,X_j,\cdots,X_l$ and the set $\{q,r,\cdots, t\}=\{1,2,\cdots,n\}\setminus\{i,j,\cdots, l\}.$ Clearly, $I_{n}(X_1: X_2: \cdots : X_n)$ is   symmetry   about the subsystems.

For the quantum states in the set  $\triangle_{\mathrm{IQN}}$, it is easy to obtain the following result.

\textbf{\emph{Conclusion 1-1.}} $I_{4}(A:B:C:D)=0$ for any $\rho\in\triangle_{\mathrm{IQN}}$.

 Now we discuss the effect of the local channels on the four-partite quantum  mutual information. Suppose that there are three local channels $\Lambda_ A$, $\Lambda_B$, and $\Lambda_C$ acting on nodes \emph{A}, \emph{B}, and \emph{C} respectively. The four-partite quantum  mutual information of the quantum system is $I_{4}[(\Lambda_{A}\otimes \Lambda_{B}\otimes \Lambda_{C}\otimes I_D)\rho]$. Here $I_D$ is the identity operator on the node $D$, and $\rho\in\triangle_{\mathrm{IQN}}$.

Clearly, in this case for two particles in \emph{D} we have $S(D)=S(D_{\alpha})+S(D_{\delta})$. By considering the actions of the local  quantum channels we can get
\begin{equation}
\begin{aligned}
x_{1}=&S(A_{\alpha}A_{\beta})+S(B_{\beta}B_{\gamma})+S(C_{\gamma}C_{\delta})+S(D_{\delta})+S(D_{\alpha})+S(A_{\alpha}A_{\beta}B_{\beta}B_{\gamma}C_{\gamma}C_{\delta})\\
&+S(A_{\alpha}A_{\beta}D_{\alpha})+S(C_{\gamma}C_{\delta}D_{\delta})+S(B_{\beta}B_{\gamma}C_{\gamma}C_{\delta}D_{\delta})+S(D_{\alpha})\\
&+S(A_{\alpha}A_{\beta}B_{\beta}B_{\gamma}D_{\alpha})+S(D_{\delta}),\\
x_{2}=&S(A_{\alpha}A_{\beta}B_{\beta}B_{\gamma})+S(A_{\alpha}A_{\beta})+S(C_{\gamma}C_{\delta})+S(A_{\alpha}A_{\beta}D_{\alpha})+S(D_{\delta})\\
&+S(B_{\beta}B_{\gamma}C_{\gamma}C_{\delta})+S(B_{\beta}B_{\gamma})+S(D_{\alpha})+S(D_{\delta})+S(C_{\gamma}C_{\delta}D_{\delta})+S(D_{\alpha})\\
&+S(D_{\alpha}A_{\alpha}A_{\beta}B_{\beta}B_{\gamma}C_{\gamma}C_{\delta}D_{\delta})\notag.\\
\end{aligned}
\end{equation}
Therefore, we derive that
\begin{equation}
\begin{aligned}
& I_{4}[(\Lambda_{A}\otimes \Lambda_{B}\otimes \Lambda_{C}\otimes I_D)\rho]\\
=&x_{1}-x_{2}\\
=&S(A_{\alpha}A_{\beta}B_{\beta}B_{\gamma}C_{\gamma}C_{\delta})+S(A_{\alpha}A_{\beta}B_{\beta}B_{\gamma}D_{\alpha})+S(B_{\beta}B_{\gamma}C_{\gamma}C_{\delta}D_{\delta})\\
&- S(A_{\alpha}A_{\beta}B_{\beta}B_{\gamma})-S(B_{\beta}B_{\gamma}C_{\gamma}C_{\delta})-S(A_{\alpha}A_{\beta}B_{\beta}B_{\gamma}C_{\gamma}C_{\delta}D_{\delta}D_{\alpha})\notag.\\
=&S(ABC)+S(ABD_{\alpha})+S(BCD_{\delta})- S(AB)-S(BC)-S(ABC D_\delta D_\alpha)\notag.
\end{aligned}
\end{equation}
Hence, generally speaking, one can not determine whether $I_4$ is larger than zero  or not for the quantum state   $(\Lambda_{A}\otimes \Lambda_{B}\otimes \Lambda_{C}\otimes I_D)\rho$.

However, for two adjacent local channels \cite{24,25,26}, the four-partite quantum  mutual information of the quantum system becomes $I_{4}[(\Lambda_{A}\otimes \Lambda_{B}\otimes I_C\otimes I_D)\rho]$. For this case, one has
\begin{equation}
\begin{aligned}
&I_{4}[(\Lambda_{A}\otimes \Lambda_{B}\otimes I_C\otimes I_D)\rho]\\
=&S(A_{\alpha}A_{\beta}B_{\beta}B_{\gamma}C_{\gamma})+S(A_{\alpha}A_{\beta}B_{\beta}B_{\gamma}D_{\alpha})- S(A_{\alpha}A_{\beta}B_{\beta}B_{\gamma})-S(A_{\alpha}A_{\beta}B_{\beta}B_{\gamma}C_{\gamma}D_{\alpha})\\
=&S(ABC_{\gamma})+S(ABD_{\alpha})- S(AB)-S(ABC_{\gamma}D_{\alpha}).\nonumber
\end{aligned}
\end{equation}
By using the strong subadditivity inequality
\begin{equation}\label{8}
S(XY)+S(XZ)\geq S(X)+S(XYZ)
\end{equation}
for  three quantum systems $X,Y,Z,$ it is easy to  obtain that the  four-partite quantum  mutual information of quantum system satisfies $I_{4}[(\Lambda_{A}\otimes\Lambda_{B}\otimes I_C\otimes I_D)\rho]\geq 0$.

For two non-adjacent local channels, the four-partite quantum  mutual information of the quantum system turns out to be $I_{4}[(\Lambda_{A}\otimes I_B\otimes\Lambda_{C}\otimes I_D)\rho]$. In this case one can deduce
\begin{equation}
I_{4}[(\Lambda_{A}\otimes I_{B}\otimes \Lambda_{C}\otimes I_D)\rho]=S(AB_{\beta}D_{\alpha})+S(B_{\gamma}CD_{\delta})-S(AB_{\beta}D_\alpha B_\gamma C  D_\delta).
\end{equation}
By the subadditivity inequality
\begin{equation}\label{8}
S(X)+S(Y)\geq S(XY)
\end{equation}
for  two quantum systems $X,Y$, we arrive at that
 the four-partite quantum  mutual information of the quantum system $I_{4}[(\Lambda_{A}\otimes I_B\otimes\Lambda_{C}\otimes I_D)\rho]$ is still large than zero. Combining the above two cases we have the following result.

\textbf{\emph{Conclusion 1-2.}} For any pair of local channels  one finds that the four-partite quantum  mutual information for the quantum states obtained by acting the pair of the  local quantum channels on the quantum state   in $\triangle _{\mathrm{IQN}}$  cannot be rendered negative.

Although in general  we can not decide  $I_{4}[(\Lambda_{A}\otimes \Lambda_{B}\otimes \Lambda_{C}\otimes I_D)\rho]$ is large or not than zero, however, we have the following conclusion.

 \textbf{\emph{Conclusion 1-3.}} The four-partite quantum  mutual information $I_{4}[(\Lambda_{A}\otimes \Lambda_{B}\otimes \Lambda_{C}\otimes I_D)\rho]$ are bounded as
\begin{eqnarray}
& &2I_2(A:D)+2I_2(B:D)+2I_2(C:D)-[I_2(AB:D)+I_2(BC:D)+I_2(AC:D)]\\\nonumber
&\leq &I_{4}[(\Lambda_{A}\otimes \Lambda_{B}\otimes \Lambda_{C}\otimes I_D)\rho]\\\nonumber
&\leq &I_2(ABC:D)-[I_2(A:D)+I_2(B:D)+I_2(C:D)]
\end{eqnarray}
and
\begin{eqnarray}
2I_2(A:D)+2I_2(B:D)+2I_2(C:D)-[I_2(AB:D)+I_2(BC:D)+I_2(AC:D)]\leq 0,
\end{eqnarray}
\begin{eqnarray}
I_2(ABC:D)-[I_2(A:D)+I_2(B:D)+I_2(C:D)]\geq 0.
\end{eqnarray}
Here  $\rho$ is a quantum state in the set $\Delta_{\mathrm{IQN}}$.

 For detailed proof of above conclusion please refer to Appendix A.

As there exists quantum entanglement in Fig. 1, so the entanglement measure $\varepsilon[\sigma]$ should be introduced. The entanglement measure should satisfy  three requirements: (1) if the quantum state is separable, the entanglement measure will disappear; (2) under local operation and classical communication, the average entanglement measure does not increase; (3) under local unitary transformation, the entanglement measure does not change. For present situation, we also require that the entanglement measure is additive on tensor products,  that is
\begin{equation}\label{4}
\begin{aligned}
\varepsilon[\sigma_{1}\otimes \sigma_{2}\otimes \sigma_{3}\otimes \sigma_{4}]=\varepsilon[\sigma_{1}]+\varepsilon[\sigma_{2}]+\varepsilon[\sigma_{3}]+\varepsilon[\sigma_{4}].
\end{aligned}
\end{equation}

Furthermore, we require the entanglement measure satisfies the monogamy relation \cite{23},
\begin{equation}\label{5}
\begin{aligned}
\varepsilon_{T|X}[\sigma_{TX}]+\varepsilon_{T|Y}[\sigma_{TY}]+\varepsilon_{T|Z}[\sigma_{TZ}]\leq\varepsilon_{T|XYZ}[\sigma_{TXYZ}],\\
\end{aligned}
\end{equation}
where $\sigma_{TX}=\mathrm{tr}_{YZ}\sigma_{TXYZ}$, $\sigma_{TY}=\mathrm{tr}_{XZ}\sigma_{TXYZ}$. Clearly, not arbitrary entanglement measure meets the above requirements, however the squashed entanglement does \cite{35, 36}.

 For the quantum state (1), as the local unitary matrices do not change the entanglement between the parties, we  have
\begin{equation}
\varepsilon_{A|BCD}=\varepsilon_{A_{\alpha}A_{\beta}|B_{\beta}B_{\gamma}C_{\gamma}C_{\delta}D_{\delta}D_{\alpha}}=\varepsilon_{A_{\alpha}|D_{\alpha}}+\varepsilon_{A_{\beta}|B_{\beta}}
=\varepsilon_{A|D}+\varepsilon_{A|B},
\end{equation}
where $A_{\alpha}$ stands for a particle that $A$ receives from  $\rho_{\alpha}$, similarly for others. Therefore for the entanglement measure $\varepsilon[\sigma]$ satisfying the above requirements,  we have the following conclusion.

\textbf{\emph{Conclusion 1-4.}}  For any $\rho\in \triangle_{\mathrm{IQN}}$, $\varepsilon_{T|XYZ}[\rho]=\varepsilon_{T|X}[\mathrm{tr}_{YZ}\rho]+\varepsilon_{T|Z}[\mathrm{tr}_{XY}\rho]$ holds for all bipartitions, such as $A|BCD$, $B|ACD$, $C|ABD$, and $D|ABC$.

After that  we discuss  the ranks of the global state (1) and  its marginals. Because the unitary matrices  $U_{A}$, $U_{B}$, $U_{C}$, and $U_{D}$ don't change the rank of the global state, for all $\rho\in\Delta_{\mathrm{IQN}}$, we have the rank of quantum state $\rho$
$$\mathrm{rk}(\rho)=\mathrm{rk}(\rho_{\alpha})\mathrm{rk}(\rho_{\beta})\mathrm{rk}(\rho_{\gamma})\mathrm{rk}(\rho_{\delta})=r_{\alpha}r_{\beta}r_{\gamma}r_{\delta},$$
where $r_{\alpha}=\mathrm{rk}(\rho_\alpha)$ is the rank of quantum state $\rho_\alpha$, similarly for others.  The rank of the local reduced state is
\begin{equation}
\mathrm{rk}(\mathrm{tr}_{BCD}\rho)=\mathrm{rk}(\mathrm{tr}_{D}\rho_{\alpha})\mathrm{rk}(\mathrm{tr}_{B}\rho_{\beta})\notag=r_{\alpha}^{A}r_{\beta}^{A}\notag,
\end{equation}
the rank of the two-partite reduced state is
\begin{equation}
\mathrm{rk}(\mathrm{tr}_{CD}\rho)=\mathrm{rk}(\mathrm{tr}_{D}\rho_{\alpha})\mathrm{rk}(\mathrm{tr}_{CD}\rho_{\beta})\mathrm{rk}(\mathrm{tr}_{C}\rho_{\gamma})
=r_{\alpha}^{A}r_{\beta}r_{\gamma}^{B}\notag,
\end{equation}
the rank of the three-partite reduced state is
\begin{equation}
\mathrm{rk}(\mathrm{tr}_{D}\rho)=\mathrm{rk}(\mathrm{tr}_{D}\rho_{\alpha})\mathrm{rk}(\rho_{\beta})\mathrm{rk}(\rho_{\gamma})\mathrm{rk}(\mathrm{tr}_{D}\rho_{\delta})
=r_{\alpha}^{A}r_{\beta}r_{\gamma}r_{\delta}^{C}\notag.
\end{equation}
Here $r_{\alpha}^{A}=\mathrm{rk}(\mathrm{tr}_{D}\rho_\alpha)$, similarly for others.

Then  we arrive at
\begin{equation}\label{188}
\begin{array}{llllll}
\mathrm{rk}(\rho)=r_{\alpha}r_{\beta}r_{\gamma}r_{\delta}, & & & \\
\mathrm{rk}(\mathrm{tr}_{BCD}\rho)=r_{\alpha}^{A}r_{\beta}^{A}, &\quad \mathrm{rk}(\mathrm{tr}_{CDA}\rho)=r_{\beta}^{B}r_{\gamma}^{B},& \quad\mathrm{rk}(\mathrm{tr}_{DAB}\rho)=r_{\gamma}^{C}r_{\delta}^{C}, &\quad \mathrm{rk}(\mathrm{tr}_{ABC}\rho)=r_{\alpha}^{D}r_{\delta}^{D},\\
\mathrm{rk}(\mathrm{tr}_{CD}\rho)=r_{\alpha}^{A}r_{\beta}r_{\gamma}^{B},& \quad \mathrm{rk}(\mathrm{tr}_{CA}\rho)=r_{\alpha}^{D}r_{\beta}^{B}r_{\gamma}^{B}r_{\delta}^{D},&
\quad\mathrm{rk}(\mathrm{tr}_{CB}\rho)=r_{\alpha}r_{\beta}^{A}r_{\delta}^{D},&\\
\mathrm{rk}(\mathrm{tr}_{DA}\rho)=r_{\beta}^{B}r_{\gamma}r_{\delta}^{C},&\quad \mathrm{rk}(\mathrm{tr}_{DB}\rho)=r_{\alpha}^{A}r_{\beta}^{A}r_{\gamma}^{C}r_{\delta}^{C}, &\quad \mathrm{rk}(\mathrm{tr}_{AB}\rho)=r_{\alpha}^{D}r_{\gamma}^{C}r_{\delta},&\\
\mathrm{rk}(\mathrm{tr}_{A}\rho)=r_{\alpha}^{D}r_{\beta}^{B}r_{\gamma}r_{\delta}, &\quad \mathrm{rk}(\mathrm{tr}_{B}\rho)=r_{\alpha}r_{\beta}^{A}r_{\gamma}^{C}r_{\delta},&\quad
\mathrm{rk}(\mathrm{tr}_{C}\rho)=r_{\alpha}r_{\beta}r_{\gamma}^{B}r_{\delta}^{D}, &\quad \mathrm{rk}(\mathrm{tr}_{D}\rho)=r_{\alpha}^{A}r_{\beta}r_{\gamma}r_{\delta}^{C}.
\end{array}
\end{equation}
If the Hilbert space of  every particle is $\emph{d}$-dimensional, we have $r_{\alpha}$, $r_{\beta}$, $r_{\gamma}$, $r_{\delta}$ $\in[1,d^{2}]$, $r_{\alpha}^{A}$, $r_{\alpha}^{D}$, $r_{\beta}^{A}$, $r_{\beta}^{B}$, $r_{\gamma}^{B}$, $r_{\gamma}^{C}$, $r_{\delta}^{C}$, $r_{\delta}^{D}$ $\in[1,d]$. Hence one has the conclusion as follows.

\textbf{\emph{Conclusion 1-5.}} For $\rho \in\triangle_{\mathrm{IQN}}$ there exist integers $r_{\alpha}$, $r_{\beta}$, $r_{\gamma}$, $r_{\delta}$$\in[1,d^{2}]$, $r_{\alpha}^{A}$, $r_{\alpha}^{D}$, $r_{\beta}^{A}$, $r_{\beta}^{B}$, $r_{\gamma}^{B}$, $r_{\gamma}^{C}$, $r_{\delta}^{C}$, $r_{\delta}^{D}$$\in[1,d]$, which satisfy rank relationship Eq. (\ref{188}).

Next we discuss the scenario where the four nodes and the four sources are classically correlated \cite{18}.
As shown in Fig. 1(b), we assume that every source and every node are  correlated by sharing the random variable $\lambda$ \cite{18}. In this case, the quantum states are called classically  correlated states, which  read
\begin{equation}\label{}
\begin{aligned}
\rho=&\int \texttt{d}\lambda p(\lambda)(U_{A}(\lambda)\otimes U_{B}(\lambda)\otimes U_{C}(\lambda)\otimes U_{D}(\lambda))(\rho_{\alpha}(\lambda)\otimes\rho_{\beta}(\lambda)\otimes\rho_{\gamma}(\lambda)\otimes\rho_{\delta}(\lambda)\\
&(U_{A}^{\dag}(\lambda)\otimes U_{B}^{\dag}(\lambda)\otimes U_{C}^{\dag}(\lambda)\otimes U_{D}^{\dag}(\lambda)).
\end{aligned}
\end{equation}
Here $p(\lambda)$ is a probability distribution function of the random variable $\lambda$.
We use CQN to denote this classically correlated quadrangular  network,  $\Delta_{\mathrm{CQN}}$ to represent the set of quantum states that can be prepared in CQN. In other words, a quantum state $\rho \in \Delta_{\mathrm{CQN}}$ can be written as $\rho=\sum_{\lambda}p_{\lambda}\rho_{\lambda}$, where $\rho_{\lambda} \in \Delta_{\mathrm{IQN}}$, $p_{\lambda}$ is a probability distribution.

Based on the properties of the ranks of quantum states, we will verify that there are no four-qubit genuine  multipartite entangled states in the set  $\Delta_{\mathrm{CQN}}$.  Here   four-qubit genuine  multipartite entangled state refers to it can be embedded in larger dimensional systems. Evidently, if a  four-qubit genuine multipartite entangled state  is pure, then the rank of the global state $\rho$ is equal to 1 \cite{27,28,29} and it is entangled along every bipartition. Considering the Schmidt decomposition of the four-qubit genuine multipartite entangled state, one has that all single node reduced state has rank 2. However, as a matter of fact, when each source in the IQN   is entangled among two partitions. According to the Schmidt decomposition, the rank of the reduced state of a particle in each source  is  2. Therefore, in IQN as shown in Fig. 1(a), with sources which  prepare 2-qubit entangled state, the local rank of the reduced state at each node is 4.  So it is impossible to prepare  four-qubit genuine multipartite entangled states in  IQN. Similarly, we can show that the statement holds for other cases of resources. Furthermore, because  four-qubit genuine  entangled states  need  pure four-qubit genuine multipartite entangled  states \cite{30}, so it is also impossible to prepare four-qubit genuine  multipartite entangled states in  CQN. Hence we arrive at the following conclusion.

\textbf{\emph{Conclusion 1-6.}} No four-qubit genuine multipartite entangled state can be prepared in  IQN and CQN.

Fig. 1(c) shows the four-partite quantum states which can not be generated in IQN and CQN.

\section{Four nodes in the space}

Clearly, four nodes in the space form a tetrahedron.  By the assumption of no intersection of sources, in this case there are two different quantum network structures: one  is the triangular cone network structure 1  with six two-partite   sources as shown in Fig. 2(a); the other is the triangular cone network structure 2 with four three-partite sources as illustrated in Fig. 3(a).

We first study the triangular cone network structure 1. It has four nodes, namely \emph{A}, \emph{B}, \emph{C}, and \emph{D}. These four nodes are formed by pairs of two particles in each of the six sources $\rho_{\alpha}$, $\rho_{\beta}$, $\rho_{\gamma}$, $\rho_{\delta}$, $\rho_{\theta}$, and $\rho_{\tau}$.  Two particles in each source are in two-partite entangled states.  $\rho_{\alpha}$,  $\rho_{\beta}$,  $\rho_{\gamma}$, $\rho_{\delta}$,  $\rho_{\theta}$ and  $\rho_{\tau}$ are shared by $\{A, B\}$,   $\{B, D\}$,  $\{A, D\}$,  $\{A, C\}$, $\{C, D\}$,  $\{B, C\}$, respectively.  Four nodes do not share common information. Each of  four nodes \emph{A}, \emph{B}, \emph{C}, and \emph{D}  receives  three particles, for example, ${A}$ receives three particles from sources $\rho_{\alpha}$, $\rho_{\gamma}$, and $\rho_{\delta}$, similarly for others. One can apply a local unitary matrix to the three received particles. Four local unitary matrices are denoted by $U_{A}$, $U_{B}$, $U_{C}$, and $U_{D}$ respectively.  The global  quantum state in an  independent triangular cone network structure 1 (ITCN1) is denoted by $\rho$. We use $\Delta_{\mathrm{ITCN1}}$ to represent the set of quantum states that can be prepared in ITCN1.

\begin{figure}[h]
\centering
\includegraphics[height=0.33\textwidth]{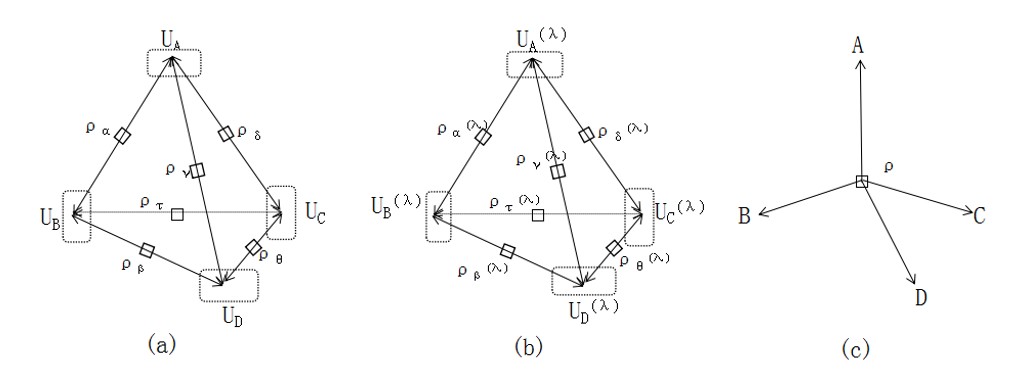}
\caption{Triangular cone network structure 1. (a) Independent triangular cone network structure 1 (ITCN1), where each of these sources is a two-partite entangled state. (b) Classically correlated  triangular cone network structure 1 (CTCN1), where each source and each node are classically correlated by sharing the random variable $\lambda$.  (c) The four-partite quantum states which can not be prepared in  ITCN1 and CTCN1.
\label{FIG. 2.}}
\end{figure}

Apparently, the quantum states prepared in ITCN1 read
\begin{equation}\label{66}
\rho=(U_{A}\otimes U_{B}\otimes U_{C}\otimes U_{D})(\rho_{\alpha}\otimes\rho_{\beta}\otimes\rho_{\gamma}\otimes\rho_{\delta}\otimes\rho_{\theta}\otimes\rho_{\tau})(U_{A}^{\dag}\otimes U_{B}^{\dag}\otimes U_{C}^{\dag}\otimes U_{D}^{\dag}).
\end{equation}

By using that entropy is invariant under unitary transformation, and for the tensor product state, entropy has additivity, we  have the entropy of  the quantum state (\ref{66})  $$S(\rho)=S(\rho_{\alpha})+S(\rho_{\beta})+S(\rho_{\gamma})+S(\rho_{\delta})+S(\rho_{\theta})+S(\rho_{\tau}).$$
The three-partite entropy, the two-partite entropy,  the one-partite entropy  are extended to
$$S(ABC)=S(\rho_{\alpha})+S(\mathrm{tr}_{D}\rho_{\beta})+S(\mathrm{tr}_{D}\rho_{\gamma})+S (\rho_{\delta})+S(\mathrm{tr}_{D}\rho_{\theta})+S(\rho_{\tau}),$$
 $$S(AB)=S( \rho_{\alpha})+S(\mathrm{tr}_{D}\rho_{\beta})+S(\mathrm{tr}_{D}\rho_{\gamma})+S(\mathrm{tr}_{C}\rho_{\delta})+S(\mathrm{tr}_{C}\rho_{\tau}),$$
 $$S(A)=S(\mathrm{tr}_{B}\rho_{\alpha})+S(\mathrm{tr}_{D}\rho_{\gamma})+S(\mathrm{tr}_{C}\rho_{\delta}),$$
 respectively, and similarly for the other cases. It is easy to  obtain that $I_{4}(A:B:C:D)=0$. Hence we have the following result.

\textbf{\emph{Conclusion 2-1-1.}}  For any $\rho\in\triangle_{\mathrm{ITCN1}}$, the four-partite quantum mutual information $I_{4}(A:B:C:D)$ of quantum state $\rho$ is zero.

Now we discuss entanglement measure. We require the entanglement measure $\varepsilon$ satisfies the requirements stated in section II. It is easy to deduce
\begin{equation}
\begin{array}{ll}
\varepsilon_{A|BCD}&=\varepsilon_{A_{\alpha}A_{\gamma}A_{\delta}|B_{\alpha}B_{\beta}B_{\tau}C_{\delta}C_{\theta}C_{\tau}D_{\beta}D_{\gamma}D_{\theta}}=\varepsilon_{A_{\alpha}|B_{\alpha}}+\varepsilon_{A_{\gamma}|D_{\gamma}}+\varepsilon_{A_{\delta}|C_{\delta}}
=\varepsilon_{A|B}+\varepsilon_{A|D}+\varepsilon_{A|C}.\\
\end{array}
\end{equation}
So the following result holds.

\textbf{\emph{Conclusion 2-1-2.}} For any $\rho\in \triangle_{\mathrm{ITCN1}}$, we have that $\varepsilon_{T|XYZ}[\rho]=\varepsilon_{T|X}[\mathrm{tr}_{YZ}\rho]+\varepsilon_{T|Y}[\mathrm{tr}_{XZ}\rho]+\varepsilon_{T|Z}[\mathrm{tr}_{XY}\rho]$,  for all the bipartitions, such as $A|BCD$, $B|ACD$, $C|ABD$, and $D|ABC$.

Later on  we turn our attention to the rank of the quantum states. Because the unitary matrices  $U_{A}$, $U_{B}$, $U_{C}$, and $U_{D}$ don't change the rank of the whole quantum system, for all $\rho\in\Delta_{\mathrm{ITCN1}}$, we have the rank of the quantum state $\rho$, $$\mathrm{rk}(\rho)=\mathrm{rk}(\rho_{\alpha})\mathrm{rk}(\rho_{\beta})\mathrm{rk}(\rho_{\gamma})\mathrm{rk}(\rho_{\delta})\mathrm{rk}(\rho_{\theta})\mathrm{rk}(\rho_{\tau})=r_{\alpha}r_{\beta}r_{\gamma}r_{\delta}r_{\theta}r_{\tau},$$
the rank of the local reduced state  $\mathrm{tr}_{BCD}\rho$ is
\begin{equation}
\mathrm{rk}(\mathrm{tr}_{BCD}\rho)=\mathrm{rk}(\mathrm{tr}_{B}\rho_{\alpha})\mathrm{rk}(\mathrm{tr}_{D}\rho_{\gamma})\mathrm{rk}(\mathrm{tr}_{C}\rho_{\delta})=r_{\alpha}^{A}r_{\gamma}^{A}r_{\delta}^{A}\notag,\\
\end{equation}
the rank of the two-partite reduced state $\mathrm{tr}_{CD}\rho$ is
\begin{equation}
\mathrm{rk}(\mathrm{tr}_{CD}\rho)=\mathrm{rk}(\rho_{\alpha})\mathrm{rk}(\mathrm{tr}_{D}\rho_{\beta})\mathrm{rk}(\mathrm{tr}_{D}\rho_{\gamma})\mathrm{rk}(\mathrm{tr}_{C}\rho_{\delta})\mathrm{rk}(\mathrm{tr}_{C}\rho_{\tau})=r_{\alpha}r_{\beta}^{B}r_{\gamma}^{A}r_{\delta}^{A}r_{\tau}^{B}\notag,\\
\end{equation}
the rank of the three-partite  reduced state $\mathrm{tr}_{D}\rho$ is
\begin{equation}
\mathrm{rk}(\mathrm{tr}_{D}\rho)=\mathrm{rk}(\rho_{\alpha})\mathrm{rk}(\mathrm{tr}_{D}\rho_{\beta})\mathrm{rk}(\mathrm{tr}_{D}\rho_{\gamma})\mathrm{rk}(\rho_{\delta})\mathrm{rk}(\mathrm{tr}_{D}\rho_{\theta})\mathrm{rk}(\rho_{\tau})=r_{\alpha}r_{\beta}^{B}r_{\gamma}^{A}r_{\delta}r_{\theta}^{C}r_{\tau}.\notag\\
\end{equation}
Here $r_\alpha^A=\mathrm{rk}(\mathrm{tr}_B\rho_\alpha)$ stands for the rank of the quantum state $\mathrm{tr}_B\rho_\alpha$, similarly for the others.

Therefore we have
\begin{equation}\label{r1}
\begin{array}{llll}
\mathrm{rk}(\rho)=r_{\alpha}r_{\beta}r_{\gamma}r_{\delta}r_{\theta}r_{\tau},& & &\\
\mathrm{rk}(\mathrm{tr}_{BCD}\rho)=r_{\alpha}^{A}r_{\gamma}^{A}r_{\delta}^{A}, & \quad \mathrm{rk}(\mathrm{tr}_{CDA}\rho)=r_{\alpha}^{B}r_{\beta}^{B}r_{\tau}^{B}, &\quad\mathrm{rk}(\mathrm{tr}_{DAB}\rho)=r_{\delta}^{C}r_{\theta}^{C}r_{\tau}^{C}, & \quad \mathrm{rk}(\mathrm{tr}_{ABC}\rho)=r_{\beta}^{D}r_{\gamma}^{D}r_{\theta}^{D},\\
\mathrm{rk}(\mathrm{tr}_{CD}\rho)=r_{\alpha}r_{\beta}^{B}r_{\gamma}^{A}r_{\delta}^{A}r_{\tau}^{B}, &\quad \mathrm{rk}(\mathrm{tr}_{CA}\rho)=r_{\alpha}^{B}r_{\beta}r_{\gamma}^{D}r_{\theta}^{D}r_{\tau}^{B}, &
\quad\mathrm{rk}(\mathrm{tr}_{CB}\rho)=r_{\alpha}^{A}r_{\beta}^{D}r_{\gamma}r_{\delta}^{A}r_{\theta}^{D},& \\
\mathrm{rk}(\mathrm{tr}_{DA}\rho)=r_{\alpha}^{B}r_{\beta}^{B}r_{\delta}^{C}r_{\theta}^{C}r_{\tau},&
\quad\mathrm{rk}(\mathrm{tr}_{DB}\rho)=r_{\alpha}^{A}r_{\gamma}^{A}r_{\delta}r_{\theta}^{C}r_{\tau}^{C}, & \quad\mathrm{rk}(\mathrm{tr}_{AB}\rho)=r_{\beta}^{D}r_{\gamma}^{D}r_{\delta}^{C}r_{\theta}r_{\tau}^{C},& \\
\mathrm{rk}(\mathrm{tr}_{A}\rho)=r_{\alpha}^{B}r_{\beta}r_{\gamma}^{D}r_{\delta}^{C}r_{\theta}r_{\tau}, &\quad \mathrm{rk}(\mathrm{tr}_{B}\rho)=r_{\alpha}^{A}r_{\beta}^{D}r_{\gamma}r_{\delta}r_{\theta}r_{\tau}^{C}, &
\quad\mathrm{rk}(\mathrm{tr}_{C}\rho)=r_{\alpha}r_{\beta}r_{\gamma}r_{\delta}^{A}r_{\theta}^{D}r_{\tau}^{B}, &\quad \mathrm{rk}(\mathrm{tr}_{D}\rho)=r_{\alpha}r_{\beta}^{B}r_{\gamma}^{A}r_{\delta}r_{\theta}^{C}r_{\tau}.
\end{array}
\end{equation}

Suppose that the Hilbert space of single particle is $\emph{d}$-dimensional. So $r_{\alpha}$, $r_{\beta}$, $r_{\gamma}$, $r_{\delta}$, $r_{\theta}$, $r_{\tau}$$\in[1,d^{2}]$, $r_{\alpha}^{A}$, $r_{\alpha}^{B}$, $r_{\beta}^{B}$, $r_{\beta}^{D}$, $r_{\gamma}^{A}$, $r_{\gamma}^{D}$, $r_{\delta}^{A}$, $r_{\delta}^{C}$, $r_{\theta}^{C}$, $r_{\theta}^{D}$, $r_{\tau}^{B}$, $r_{\tau}^{C}$$\in[1,d]$. Thus we arrive at the following conclusion.

\textbf{\emph{Conclusion 2-1-3. }}For $\rho \in\triangle_{\mathrm{ITCN1}}$ there exist integers $r_{\alpha}$, $r_{\beta}$, $r_{\gamma}$, $r_{\delta}$, $r_{\theta}$, $r_{\tau}$$\in[1,d^{2}]$, $r_{\alpha}^{A}$, $r_{\alpha}^{B}$, $r_{\beta}^{B}$, $r_{\beta}^{D}$, $r_{\gamma}^{A}$, $r_{\gamma}^{D}$, $r_{\delta}^{A}$, $r_{\delta}^{C}$, $r_{\theta}^{C}$, $r_{\theta}^{D}$, $r_{\tau}^{B}$, $r_{\tau}^{C}$$\in[1,d]$,  which satisfy Eq. (\ref{r1}).

After that  we invistegate the situation  where the four nodes and the six sources are classically correlated \cite{18}. The classically correlated refers to  that every source and every node are correlated by sharing the random variable $\lambda$. The classically correlated triangular cone network structure 1 is shown in Fig. 2(b). We use CTCN1 to denote this classically correlated triangular cone network, $\Delta_{\mathrm{CTCN1}}$ to represent the set of quantum states that can be prepared in CTCN1. A quantum state $\rho \in \Delta_{\mathrm{CTCN1}}$ can be written as $\rho=\sum_{\lambda}p_{\lambda}\rho_{\lambda}$, where $\rho_{\lambda} \in \Delta_{\mathrm{ITCN1}}$, $p_{\lambda}$ is a probability distribution.

Next we will use the rank relationship to show that there exists no four-qubit genuine multipartite entangled state in the set $\Delta_{\mathrm{CTCN1}}$. First we consider the case that the four-qubit genuine multipartite entangled state is a pure state. Clearly, in this case,  the rank of the four-qubit genuine multipartite entangled state is equal to 1 and it is entangled along every bipartition. By the Schmidt decomposition of the four-qubit genuine  entangled state, we know  that all single node reduced state has rank 2. However, in fact, when each source in the set $\{\rho_\alpha, \rho_\beta, \rho_\gamma, \rho_\theta, \rho_\tau, \rho_\delta\}$ is entangled among two particles. According to the Schmidt decomposition, the rank of the reduced state of a particle in each source is 2, if source is in 2-qubit entangled state.  For the network structure ITCN1 with six sources as shown in Fig. 2(a),  the local rank of the reduced states of each node is 8. Therefore, it is impossible to prepare pure four-qubit genuine multipartite entangled   state in  ITCN1. Similarly, one can verify that the claim is true for the other cases of resources.  Moreover, as  four-qubit genuine multipartite entangled states need pure four-qubit genuine multipartite  entangled states. So one can not prepare four-qubit genuine multipartite entangled states in  CTCN1 also. Hence, we obtain the following result.

\textbf{\emph{Conclusion 2-1-4.}} No four-qubit genuine  multipartite entangled state can be prepared in  ITCN1 and CTCN1.

The four-partite e quantum states which can not be generated  in ITCN1 and CTCN1 is illustrated in Fig. 2(c).

Now let us consider the independent triangular cone network structure 2 (ITCN2).  As shown in Fig. 3(a), it has four nodes, namely \emph{A}, \emph{B}, \emph{C}, and \emph{D}. These four nodes are formed by pairs of three particles in each of the four sources $\rho_{\alpha}$, $\rho_{\beta}$, $\rho_{\gamma}$, and $\rho_{\delta}$.  Each source produces entangled three-partite quantum states.  $\rho_{\alpha}$ is shared by \emph{A}, \emph{B}, and \emph{D},  $\rho_{\beta}$ is shared by \emph{A}, \emph{B}, and \emph{C},  $\rho_{\gamma}$ is shared by \emph{A}, \emph{C}, and \emph{D}, $\rho_{\delta}$ is shared by \emph{B}, \emph{C}, and \emph{D}. We assume that four nodes do not share common information.   \emph{A} receives three particles from source $\rho_{\alpha}$, $\rho_{\gamma}$, and $\rho_{\gamma}$, \emph{B} receives three particles from source $\rho_{\alpha}$, $\rho_{\beta}$, and $\rho_{\delta}$, \emph{C} receives three particles from source $\rho_{\beta}$, $\rho_{\gamma}$, and $\rho_{\delta}$, \emph{D} receives three particles from source $\rho_{\alpha}$, $\rho_{\gamma}$, and $\rho_{\delta}$. Each node can be applied a local unitary matrix to the three received particles. Four local unitary matrices are denoted by $U_{A}$, $U_{B}$, $U_{C}$, and $U_{D}$.  We use $\rho$ to express the  quantum state in ITCN2 and  utilize $\Delta_{\mathrm{ITCN2}}$ to denote the set of quantum states that can be produced in ITCN2.

\begin{figure}[h]
\centering
\includegraphics[height=0.33\textwidth]{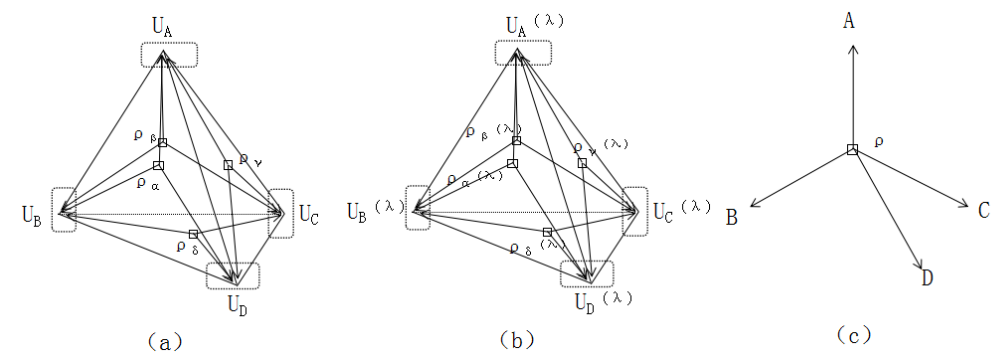}
\caption{Triangular cone network structure 2. (a) Independent triangular cone network structure 2 (ITCN2), where each of these sources is a three-partite entangled state. (b) Classically correlated triangular cone network structure 2 (CTCN2), where each source and each node are classically correlated by sharing the random variable $\lambda$.  (c) Four-partite quantum states which  can not be produced  in  ITCN2 and CTCN2.
\label{FIG. 3 .}}
\end{figure}

Obviously, the quantum states $\rho$ can be expressed as
\begin{equation}\label{222}
\rho=(U_{A}\otimes U_{B}\otimes U_{C}\otimes U_{D})(\rho_{\alpha}\otimes\rho_{\beta}\otimes\rho_{\gamma}\otimes\rho_{\delta})(U_{A}^{\dag}\otimes U_{B}^{\dag}\otimes U_{C}^{\dag}\otimes U_{D}^{\dag}).
\end{equation}
 Because at each node the three particles  are coming from  different sources, so the unitary matrix at one node  cannot act on the global  quantum states $\rho_{\alpha}$, $\rho_{\beta}$, $\rho_{\gamma}$, and $\rho_{\delta}$ at same time.

Evidently, we have that the entropy of quantum state  (\ref{222})
$$S(\rho)=S(\rho_{\alpha})+S(\rho_{\beta})+S(\rho_{\gamma})+S(\rho_{\delta}).$$
The three-partite entropy is extended to
$$S(ABC)=S(\mathrm{tr}_{D}\rho_{\alpha})+S(\rho_{\beta})+S(\mathrm{tr}_{D}\rho_{\gamma})+S (\mathrm{tr}_{D}\rho_{\delta}),$$
the two-partite entropy is extended to $$S(AB)=S( \mathrm{tr}_{D}\rho_{\alpha})+S(\mathrm{tr}_{C}\rho_{\beta})+S(\mathrm{tr}_{CD}\rho_{\gamma})+S(\mathrm{tr}_{CD}\rho_{\delta}),$$
 the one-partite entropy is extended to
   $$S(A)=S(\mathrm{tr}_{BD}\rho_{\alpha})+S(\mathrm{tr}_{BC}\rho_{\beta})+S(\mathrm{tr}_{CD}\rho_{\gamma}).$$
By calculating, one finds that for quantum state $\rho\in\triangle_{\mathrm{ITCN2}}$, $I_{4}(A:B:C:D)=0$. This can be restated as the following conclusion.

\textbf{\emph{Conclusion 2-2-1.}}  The four-partite quantum  mutual information $I_{4}(A:B:C:D)=0$ for any $\rho\in\triangle_{\mathrm{ITCN2}}$.

Then  we discuss the entanglement measure of the quantum states. For the entanglement measure $\varepsilon$ which satisfies the requirements stated in section II, we can obtain
\begin{equation}
\varepsilon_{A|BCD}=\varepsilon_{A_{\alpha}A_{\beta}A_{\gamma}|B_{\alpha}B_{\beta}B_{\delta}C_{\beta}C_{\gamma}C_{\delta}D_{\alpha}D_{\gamma}D_{\delta}}=\varepsilon_{A_{\alpha}|B_{\alpha}D_{\alpha}}+\varepsilon_{A_{\beta}|B_{\beta}C_{\beta}}+\varepsilon_{A_{\gamma}|C_{\gamma}D_{\gamma}}\\
=\varepsilon_{A|BD}+\varepsilon_{A|BC}+\varepsilon_{A|CD}.
\end{equation}
So we have the following result.

\textbf{\emph{Conclusion 2-2-2. }}For any $\rho\in \triangle_{\mathrm{ITCN2}}$, we have that $\varepsilon_{T|XYZ}[\rho]=\varepsilon_{T|XY}[\mathrm{tr}_{Z}\rho]+\varepsilon_{T|XZ}[\mathrm{tr}_{Y}\rho]+\varepsilon_{T|YZ}[\mathrm{tr}_{X}\rho]$, which holds for all the bipartitions, such as $A|BCD$, $B|ACD$, $C|ABD$, and $D|ABC$.

  As the unitary matrices $U_{A}$, $U_{B}$, $U_{C}$, and $U_{D}$ don't change the rank of the whole quantum state, so for  $\rho\in\Delta_{\mathrm{ITCN2}}$, we have  the rank of quantum state $\rho$
$$\mathrm{rk}(\rho)=\mathrm{rk}(\rho_{\alpha})\mathrm{rk}(\rho_{\beta})\mathrm{rk}(\rho_{\gamma})\mathrm{rk}(\rho_{\delta}).$$
 The rank of the local reduced state is
\begin{equation}
\mathrm{rk}(\mathrm{tr}_{BCD}\rho)=\mathrm{rk}(\mathrm{tr}_{BD}\rho_{\alpha})\mathrm{rk}(\mathrm{tr}_{BC}\rho_{\beta})\mathrm{rk}(\mathrm{tr}_{CD}\rho_{\gamma})=r_{\alpha}^{A}r_{\beta}^{A}r_{\gamma}^{A}\notag,\\
\end{equation}
the rank of the two-partite reduced state is
\begin{equation}
\mathrm{rk}(\mathrm{tr}_{CD}\rho)=\mathrm{rk}(\mathrm{tr}_{D}\rho_{\alpha})\mathrm{rk}(\mathrm{tr}_{C}\rho_{\beta})\mathrm{rk}(\mathrm{tr}_{CD}\rho_{\gamma})\mathrm{rk}(\mathrm{tr}_{CD}\rho_{\delta})=r_{\alpha}^{AB}r_{\beta}^{AB}r_{\gamma}^{A}r_{\delta}^{B}\notag,\\
\end{equation}
the rank of the three-partite reduced state is
\begin{equation}
\begin{split}
\mathrm{rk}(\mathrm{tr}_{D}\rho)=\mathrm{rk}(\mathrm{tr}_{D}\rho_{\alpha})\mathrm{rk}(\rho_{\beta})\mathrm{rk}(\mathrm{tr}_{D}\rho_{\gamma})\mathrm{rk}(\mathrm{tr}_{D}\rho_{\delta})=r_{\alpha}^{AB}r_{\beta}r_{\gamma}^{AC}r_{\delta}^{BC}.\notag
\end{split}
\end{equation}
Here $r_{\alpha}^{AB}=\mathrm{rk}(\mathrm{tr}_{D}\rho_{\alpha})$ is the rank of reduced quantum state $\mathrm{tr}_{D}\rho_{\alpha}$, similarly for the others. Therefore we have
\begin{equation}\label{r2}
\begin{array}{llll}
\mathrm{rk}(\rho)=r_{\alpha}r_{\beta}r_{\gamma}r_{\delta},& & &\\
\mathrm{rk}(\mathrm{tr}_{BCD}\rho)=r_{\alpha}^{A}r_{\beta}^{A}r_{\gamma}^{A}, & \quad \mathrm{rk}(\mathrm{tr}_{CDA}\rho)=r_{\alpha}^{B}r_{\beta}^{B}r_{\delta}^{B}, &
\quad\mathrm{rk}(\mathrm{tr}_{DAB}\rho)=r_{\beta}^{C}r_{\gamma}^{C}r_{\delta}^{C}, & \quad \mathrm{rk}(\mathrm{tr}_{ABC}\rho)=r_{\alpha}^{D}r_{\gamma}^{D}r_{\delta}^{D},\\
\mathrm{rk}(\mathrm{tr}_{CD}\rho)=r_{\alpha}^{AB}r_{\beta}^{AB}r_{\gamma}^{A}r_{\delta}^{B}, & \quad \mathrm{rk}(\mathrm{tr}_{CA}\rho)=r_{\alpha}^{BD}r_{\beta}^{B}r_{\gamma}^{D}r_{\delta}^{BD}, & \quad
\mathrm{rk}(\mathrm{tr}_{CB}\rho)=r_{\alpha}^{AD}r_{\beta}^{A}r_{\gamma}^{AD}r_{\delta}^{D}, &\\  \mathrm{rk}(\mathrm{tr}_{DA}\rho)=r_{\alpha}^{B}r_{\beta}^{BC}r_{\gamma}^{C}r_{\delta}^{BC}, & \quad
\mathrm{rk}(\mathrm{tr}_{DB}\rho)=r_{\alpha}^{A}r_{\beta}^{AC}r_{\gamma}^{AC}r_{\delta}^{C}, &\quad \mathrm{rk}(\mathrm{tr}_{AB}\rho)=r_{\alpha}^{D}r_{\beta}^{C}r_{\gamma}^{CD}r_{\delta}^{CD}, &\\
\mathrm{rk}(\mathrm{tr}_{A}\rho)=r_{\alpha}^{BD}r_{\beta}^{BC}r_{\gamma}^{CD}r_{\delta}, &\quad \mathrm{rk}(\mathrm{tr}_{B}\rho)=r_{\alpha}^{AD}r_{\beta}^{AC}r_{\gamma}r_{\delta}^{CD},& \quad
\mathrm{rk}(\mathrm{tr}_{C}\rho)=r_{\alpha}r_{\beta}^{AB}r_{\gamma}^{AD}r_{\delta}^{BD}, & \quad \mathrm{rk}(\mathrm{tr}_{D}\rho)=r_{\alpha}^{AB}r_{\beta}r_{\gamma}^{AC}r_{\delta}^{BC}.
\end{array}
\end{equation}

Assume that the Hilbert space of single particle is \emph{d}-dimensional. So $r_{\alpha}^{A}$, $r_{\alpha}^{B}$, $r_{\alpha}^{D}$, $r_{\beta}^{A}$, $r_{\beta}^{B}$, $r_{\beta}^{C}$, $r_{\gamma}^{A}$, $r_{\gamma}^{C}$, $r_{\gamma}^{D}$, $r_{\delta}^{B}$, $r_{\delta}^{C}$, $r_{\delta}^{D}$$\in[1,d]$, $r_{\alpha}^{AB}$, $r_{\alpha}^{AD}$, $r_{\alpha}^{BD}$, $r_{\beta}^{AB}$, $r_{\beta}^{AC}$, $r_{\beta}^{BC}$, $r_{\gamma}^{AC}$, $r_{\gamma}^{AD}$, $r_{\gamma}^{CD}$, $r_{\delta}^{BC}$, $r_{\delta}^{BD}$, $r_{\delta}^{CD}$$\in[1,d^{2}]$, and  $r_{\alpha}$, $r_{\beta}$, $r_{\gamma}$, $r_{\delta}$$\in[1,d^{3}]$.

\textbf{\emph{Conclusion 2-2-3.}} For $\rho \in\triangle_{\mathrm{ITCN2}}$ there exist integers $r_{\alpha}$, $r_{\beta}$, $r_{\gamma}$, $r_{\delta}$$\in[1,d^{3}]$, $r_{\alpha}^{AB}$, $r_{\alpha}^{AD}$, $r_{\alpha}^{BD}$, $r_{\beta}^{AB}$, $r_{\beta}^{AC}$, $r_{\beta}^{BC}$, $r_{\gamma}^{AC}$, $r_{\gamma}^{AD}$, $r_{\gamma}^{CD}$, $r_{\delta}^{BC}$, $r_{\delta}^{BD}$, $r_{\delta}^{CD}$$\in[1,d^{2}]$, $r_{\alpha}^{A}$, $r_{\alpha}^{B}$, $r_{\alpha}^{D}$, $r_{\beta}^{A}$, $r_{\beta}^{B}$, $r_{\beta}^{C}$, $r_{\gamma}^{A}$, $r_{\gamma}^{C}$, $r_{\gamma}^{D}$, $r_{\delta}^{B}$, $r_{\delta}^{C}$, $r_{\delta}^{D}$$\in[1,d]$,  which satisfy Eq. (\ref{r2}).

 Similar to the case investigated in ITCN1 we can study  the scenario where the four nodes and the four sources are classically correlated in ITCN2. If  every source and every node are correlated by sharing the random variable $\lambda$, we call the  quantum state prepared the classically correlated state as shown in Fig. 3(b). Let CTCN2  denote this classically correlated triangular cone network strcture 2, $\Delta_{\texttt{CTCN2}}$  express the set of quantum states that can be produced in CTCN2. The classically correlated  state $\rho \in \Delta_{\texttt{CTCN2}}$ reads $\rho=\sum_{\lambda}p_{\lambda}\rho_{\lambda}$, where $\rho_{\lambda} \in \Delta_{\mathrm{ITCN2}}$, $p_{\lambda}$ is a probability distribution.

 Based on  rank relationship we can  demonstrate that there exists no four-qubit genuine multipartite entangled state in the set $\Delta_{\mathrm{CTCN2}}$ also. First we assume four-qubit genuine multipartite entangled state is a pure state. It is easy to show  that all single node reduced state has rank 2. However, when each source in the set $\{\rho_\alpha, \rho_\beta, \rho_\gamma, \rho_\delta\}$ is entangled among three particles and it is entangled along every bipartition. Based on the Schmidt decomposition, the rank of the reduced state of one particle in each source is 2, if source is in 3-qubit entangled state.  So the local rank of the reduced states of each node is 8. This shows that  it is impossible to generate pure four-qubit genuine multipartite entangled   states in  ITCN2. Similarly, we can verify that the claim holds for other cases of resources.  Moreover,  as mixed four-qubit genuine multipartite entangled  states require  pure four-qubit genuine  multipartite entangled  states. So there exists no four-qubit genuine multipartite entangled state in the set $\Delta_{\mathrm{CTCN2}}$. That can be restated as the following result.

\textbf{\emph{Conclusion 2-2-4.}} No four-qubit genuine multipartite entangled state can be prepared in ITCN2 and CTCN2.

Fig.3(c) shows the four-partite  quantum states which can not be produced in  ITCN2 and CTCN2.

\section{Summary}

We investigate the quantum states that can be prepared and the quantum states that cannot be prepared  in three kinds of four-node network structures, including a four-node network structure in a plane and two four-node network structures in the space. The $n$-partite mutual information of quantum system, which satisfies the symmetry requirement, is defined. We analyzed the properties of four-partite quantum  mutual information of quantum states prepared in different network structures and the effect of the local channels on the four-partite quantum  mutual information. The entanglement properties of the quantum states prepared in the four-node quantum network structures are discussed. The constraints on the  rank of global state and the local ranks of the reduced states are deduced. We also prove that the four-qubit genuine multipartite entangled states can not be prepared classically in the four-node network structures. We hope that these results  may be useful for the further study of quantum network.

\begin{acknowledgments}
This work was supported by the National Natural Science Foundation of China under Grant No. 12071110, the Hebei Natural Science Foundation of China under Grant No. A2020205014, and funded by Science and Technology Project of Hebei Education Department under Grant Nos. ZD2020167, ZD2021066.

\end{acknowledgments}

\begin{appendix}
\section{ The proof of conclusion 1-3}

By Eq.(\ref{I3}), and according to Observation 4 of Ref.\cite{18}, for quantum state $(\Lambda_{A}\otimes \Lambda_{B}\otimes \Lambda_{C}\otimes I_D)\rho$,  one gets
\begin{equation}
I_{3}(A:B:D)=-I_2(AB:D)+I_2(A:D)+I_2(B:D)\leq 0,
\end{equation}
where $\rho\in \Delta_{\mathrm{IQN}}$,  $\Lambda_{A}$,  $\Lambda_{B}$, $\Lambda_{C}$ are three local quantum channels, $I_D$ is the identity operator. Similarly, we have
\begin{equation}
I_{3}(B:C:D)=-I_2(BC:D)+I_2(B:D)+I_2(C:D)\leq 0,
\end{equation}
\begin{equation}
I_{3}(A:C:D)=-I_2(AC:D)+I_2(A:D)+I_2(C:D)\leq 0,
\end{equation}
\begin{eqnarray}
I_{3}(AB:C:D)=-I_2(ABC:D)+I_2(AB:D)+I_2(C:D)\leq 0,
\end{eqnarray}
\begin{eqnarray}
I_{3}(BC:A:D)=-I_2(ABC:D)+I_2(BC:D)+I_2(A:D)\leq 0.
\end{eqnarray}

By  summing  the Eqs.(A1-A3), we have
\begin{equation}\label{88}
-I_2(AB:D)-I_2(AC:D)-I_2(BC:D)+2I_2(A:D)+2I_2(B:D)+2I_2(C:D)\leq 0.
\end{equation}

By  summing  the Eqs.(A1,A2, A4,A5), we have
\begin{equation}\label{99}
-I_2(ABC:D)+I_2(A:D)+I_2(B:D)+I_2(C:D)\leq 0.
\end{equation}

By the definition of four-partite quantum  mutual information  for an arbitrary quantum state sharing by nodes $A, B, C, D$ one can rewrite Eq.(\ref{3}) as
\begin{equation}\label{111}
\begin{array}{lll}
& &I_{4}(A:B:C:D)\\
&=&[S(ABC)+S(D)-S(ABCD)]+[S(A)+S(D)-S(AD)]\\
& &+[S(B)+S(D)-S(BD)]+[S(C)+S(D)-S(CD)]\\
& &-[S(AB)+S(D)-S(ABD)]-[S(AC)+S(D)-S(ACD)]\\
& &-[S(BC)+S(D)-S(BCD)]\\
&=&I_2(ABC:D)+I_2(A:D)+I_2(B:D)+I_2(C:D)\\
& &-I_2(AB:D)-I_2(AC:D)-I_2(BC:D).\\
\end{array}
\end{equation}

So, for  the four-partite quantum  mutual information of the quantum state $(\Lambda_{A}\otimes \Lambda_{B}\otimes \Lambda_{C}\otimes I_D)\rho$, by using Eqs. (\ref{88},\ref{99},\ref{111}), we have
\begin{equation}
\begin{array}{lll}
& &I_{4}[(\Lambda_{A}\otimes \Lambda_{B}\otimes \Lambda_{C}\otimes I_D)\rho]\\
&=&I_2(ABC:D)+I_2(A:D)+I_2(B:D)+I_2(C:D)\\
& &-I_2(AB:D)-I_2(AC:D)-I_2(BC:D)\\
&\leq &I_2(ABC:D)+I_2(A:D)+I_2(B:D)+I_2(C:D)\\
& &-[2I_2(A:D)+2I_2(B:D)+2I_2(C:D)]\\
&= &I_2(ABC:D)-[I_2(A:D)+I_2(B:D)+I_2(C:D)]\\
\\\end{array}
\end{equation}
and
\begin{equation}
\begin{array}{lll}
& &I_{4}[(\Lambda_{A}\otimes \Lambda_{B}\otimes \Lambda_{C}\otimes I_D)\rho]\\
&=&I_2(ABC:D)+I_2(A:D)+I_2(B:D)+I_2(C:D)\\
& &-I_2(AB:D)-I_2(AC:D)-I_2(BC:D)\\
&\geq &2I_2(A:D)+2I_2(B:D)+2I_2(C:D)\\
& &-[I_2(AB:D)+I_2(BC:D)+I_2(AC:D)].\\
\end{array}
\end{equation}
Therefore, $I_{4}[(\Lambda_{A}\otimes \Lambda_{B}\otimes \Lambda_{C}\otimes I_D)\rho]$ are bounded as follows
\begin{eqnarray}
& &2I_2(A:D)+2I_2(B:D)+2I_2(C:D)-[I_2(AB:D)+I_2(BC:D)+I_2(AC:D)]\\\nonumber
&\leq &I_{4}[(\Lambda_{A}\otimes \Lambda_{B}\otimes \Lambda_{C}\otimes I_D)\rho]\\\nonumber
&\leq &I_2(ABC:D)-[I_2(A:D)+I_2(B:D)+I_2(C:D)].
\end{eqnarray}

\end{appendix}

\end{document}